\documentclass{JHEP}
 \usepackage{epsfig}

\def\lsim{\:\raisebox{-0.5ex}{$\stackrel{\textstyle<}{\sim}$}\:}

\def\gsim{\:\raisebox{-0.5ex}{$\stackrel{\textstyle>}{\sim}$}\:}

 \title{
  Heavy charged Higgs boson production at next generation $e^\pm\gamma$
  colliders}

 \author{Shinya Kanemura\\
  Physics and Astronomy Department, Michigan State University\\
  East Lansing, MI 48824--1116, USA}

 \author{Stefano Moretti\\
  Theory Division, CERN, CH--1211 Gen\`eve 23, Switzerland}

 \author{Kosuke Odagiri\\
  Theory Group, KEK, 1--1 Oho, Tsukuba, Ibaraki 305--0801, Japan}

 \abstract{
  We assess the potential of future electron-positron linear
colliders operating in the
$e^\pm\gamma$ mode in detecting charged Higgs bosons with mass 
around and larger than
the top quark mass, using Compton back-scattered photons from laser
light. We compare the pair production mode, $e^-\gamma\to e^- H^+H^-$,
to a variety of channels involving only one charged Higgs scalar in the
final state, such as the tree-level processes 
$e^-\gamma\to \nu_e H^- \Phi^0 $ ($\Phi^0=h^0,H^0$ and $A^0$) and
$e^-\gamma\to \nu_e f\bar f H^-$ ($f=b,\tau$ and $\nu_\tau$) as well as
the loop-induced channel $e^-\gamma\to \nu_e H^-$. We show that, when the
charged Higgs boson mass is smaller than or comparable to 
half the collider energy, $\sqrt s_{ee}\gsim 2M_{H^\pm}$, single
production cross sections are of the same size
as the pair production rate,
whereas, for charged Higgs boson masses larger than $\sqrt s_{ee}/2$, 
all processes are heavily suppressed. In general, production  
cross sections of charged Higgs bosons via $e^\pm\gamma$ scatterings 
are smaller than those induced at an $e^+e^-$ collider and
the latter represents a better option to produce and analyse such 
particles.
}


 \preprint{CERN--TH/2001--064\\
           KEK--TH--757\\
           MSUHEP--01042}

\begin{document}

 \section{Motivation}

The physics case for exploiting the $\gamma\gamma$ and
$e^\pm\gamma$ beam options of future electron-positron linear colliders (LCs)
 in testing the Higgs sector of the electroweak interactions
is quite strong \cite{ginz}. The scenario that
one may well imagine as the legacy of the Large Hadron Collider
(LHC) era could be the following.
A neutral Higgs signal is detected at the CERN hadron collider, but no
other particles are found, and all measurements of the parameters
related to the new state (mass, width, couplings, etc.) are consistent
with those of both, e.g., the Standard Model (SM) Higgs boson, $\phi$,
and the lightest of the Minimal Supersymmetric Standard Model (MSSM)
Higgs bosons, $h^0$. (This scenario corresponds in the MSSM 
to the so-called  `decoupling regime', when the additional 
Higgs states, $H^0,A^0$ and $H^\pm$, are much heavier than the $h^0$.)

Since, to be optimistic, one should expect in the rather messy hadronic
environment of the LHC no more than a 10\% precision in the measurements 
of most of the Higgs boson 
couplings to ordinary matter (quarks, leptons and gauge vector 
bosons), it is reasonable to argue that one may have to wait till the advent of
a future leptonic machine in order to be able to pin down the exact nature
of the Higgs sector\footnote{The precision of mass and width measurements 
in the two collider environments can become comparable in some instances, 
at least for the SM Higgs particle \cite{droll}.}. 
In fact, at future $e^+e^-$ LCs,
operating in the energy range $\sqrt s_{ee}=500$ to 1000 GeV, the
accuracy of the same measurements is expected to improve to the level of
1\% or even less \cite{ginz,Marco1}. 
Besides, one can efficiently convert these machines 
to operate
in the $\gamma\gamma$ and $e^\pm\gamma$ modes.
By using Compton back-scattering of a few MeV laser light
\cite{backscattering}, one gets a
spectrum of high energy photons which emerge with mean
energy $E_\gamma\approx0.8 E_{e^\pm}$, typical spread $<\Delta E_\gamma >
~\approx0.07
E_\gamma$ and luminosity ${\cal L}_{\gamma\gamma/e\gamma}(x>
0.8x_{\mathrm{max}})\approx
\frac{1}{3}{\cal L}_{ee}$,
where $x\equiv \sqrt{s_{\gamma\gamma/e\gamma}}/\sqrt{s_{ee}}$
and $x_{\mathrm{max}}$ will be given in eq.~(\ref{xmax}) 
(see Ref.~\cite{ginz} for details). 

Under these circumstances, one could conceivably perform
high precision measurements
of the `${\mathrm{Higgs}}-\gamma-\gamma$'
vertex\footnote{Here, `Higgs' signifies
either $\phi$ or $h^0$.} in either $\gamma\gamma$ \cite{Melles}
or $e^\pm \gamma$ collisions
and of the `${\mathrm{Higgs}}-\gamma-Z$' one in the latter.
Deviations in their experimental determinations from the values predicted by
the SM  can be considered as a signal of 
New Physics \cite{all}\footnote{Recall
that these vertices occur at one-loop level in both models and can be 
mediated by new virtual MSSM charged (s)particles.}. From now on, we will
assume that the underlying dynamics of the Higgs sector is the one
of the MSSM and will start by summarising the Higgs discovery potential
of the $\gamma\gamma$ and $e^\pm\gamma$ options of future LCs
within this particular model.

The $\gamma\gamma$ mode can profitably be exploited 
in the search for other Higgs boson states, in addition to the 
scalar $h^0$, via the same reaction
which produces the latter\footnote{Hereafter, 
the symbol $\Phi^0$ collectively
refers to the three neutral Higgs boson states of the MSSM.}, 
$\gamma\gamma\to\Phi^0$ \cite{jack}.
Besides, the $\gamma\gamma\to H^+H^-$ production mode \cite{BCT}
of charged Higgs
scalars has  a cross section larger than
 the one of the $e^+e^-$ initiated mode \cite{Koma}.
As for the $e^\pm\gamma$ case, other than via $e^-\gamma\to e^-\Phi^0$
\cite{Eboli}, 
one can access neutral Higgs states via the processes
$e^-\gamma\to e^-   Z^0\Phi^0$,  
$e^-\gamma\to \nu_e W^-\Phi^0$ and  
$e^-\gamma\to \nu_e H^-\Phi^0$, with the latter mode
serving also the  purpose of generating charged Higgs scalars,
alongside the pair production channel 
$e^-\gamma\to e^-  H^+H^-$ \cite{io} (see also \cite{boos}).
Furthermore, the loop-induced production process 
$e^-\gamma\to \nu_e H^-$ is an interesting possibility \cite{KO},
which has been shown to yield sizable rates
for small values of $\tan\beta$, though  
difficult to detect because of the SM continuum background.
For both photonic environments, a detailed phenomenological simulation
(at hadron and detector  level),
similar to those already
carried out for the $e^+e^-$ mode, see Refs.~\cite{Marco1,Marco2},
does not exist to date.

It is the purpose of our study to further elaborate on the 
potential of future LCs operating in the $e^\pm\gamma$ mode
in detecting $H^\pm$ states,
 by looking at the case in which the mass of the charged Higgs
boson of the MSSM is not only heavy (i.e., near or above the top mass, 
$m_t$), as dictated by the mentioned decoupling scenario,
but also near or above half the centre-of-mass (CM) energy of the collider,
where the pair production modes of neutral and charged Higgs bosons
have exhausted their potential, because of phase space suppression.
In fact, for light enough $M_{H^\pm}$
values (i.e., below $m_t$), the processes
\begin{equation}\label{HpHm}
e^-\gamma\to e^-   H^+ H^-
\end{equation}
and
\begin{equation}\label{H0Hpm}
e^-\gamma\to \nu_e H^- \Phi^0 
\end{equation}
have already been proved to offer some chances in 
detecting such elusive particles \cite{io}. 
When $M_{H^\pm}\gsim m_t$, they can both still be exploited, but
the latter only when $\Phi^0\equiv H^0$ or $A^0$, the $h^0$ case being 
suppressed as in the heavy $M_{H^\pm}$ case the $h^0$ quickly
  decouples from the rest of the Higgs sector, see Fig.~7 of 
Ref.~\cite{io}. Here, in addition to the
two modes (\ref{HpHm})--(\ref{H0Hpm}), we also consider the channels
\begin{equation}\label{ffHpm}
e^-\gamma\to \nu_e f\bar f H^-,
\end{equation}
where $f=b,\tau$ or $\nu_\tau$, and
\begin{equation}\label{loop}
e^-\gamma\to \nu_e H^-.
\end{equation}
In particular, 
we intend to investigate whether the last two production modes can
adequately complement the first two, possibly providing an extended 
coverage in the charged Higgs sector of the MSSM, beyond the kinematic
threshold $\sqrt s_{ee}\approx 2M_{H^\pm}\approx M_{H^\pm}+M_{\Phi^0}$
(here, $\Phi^0=H^0,A^0$, in the decoupling regime).

We are also interested in assessing whether fundamental couplings of the 
underlying Higgs model can be better
measured through $e^\pm\gamma$ reactions than
in $e^+e^-$ processes, such as: the $(\gamma)W^\pm H^\mp \Phi^0$
vertices, via (\ref{H0Hpm}), the Yukawa couplings to top and bottom
quarks of both neutral and charged Higgs states, in
(\ref{ffHpm}), the `form factors' of the 
 vertex $\gamma W^\pm H^\mp$, via (\ref{loop}). 

Our present effort is meant to complement the one carried out 
in Refs.~\cite{epem} (see also \cite{talk}) and \cite{gamgam} for the case
of $e^+e^-$ and $\gamma\gamma$ collisions, respectively,
hence providing a complete overview of the feasibility of detecting
heavy charged Higgs states in a future LC environment. 

Notice that processes (\ref{HpHm})--(\ref{ffHpm}) all occur at tree level,
whereas (\ref{loop}) takes  place at one loop as
the $\gamma W^\pm H^\mp$ vertex is forbidden at tree level
because of gauge invariance. 
The Feynman graphs corresponding to the above reactions are shown in
Figs.~\ref{feynman_graphs_HpHm}--\ref{feynman_graphs_loop}, respectively.
Also notice that, in order to avoid double counting process (\ref{H0Hpm}), we 
have not included in the simulation of (\ref{ffHpm}) final states
the contribution of graphs proceeding via  intermediate $\nu_e H^-\Phi^{0}$
stages (i.e., the diagrams in Fig.~\ref{H0Hpm}, followed by
 $\Phi^0\to b\bar b$).

 \section{Production cross sections}

We have computed the production cross sections for
$e^-\gamma\to e^-   H^+ H^-$ and $e^-\gamma\to \nu_e H^-\Phi^0 $
by using the program originally developed in Ref.~\cite{io}; the
$e^-\gamma\to \nu_e f\bar f H^-$ reactions were simulated by producing
a totally new code, based on helicity amplitudes \cite{helicity_HZ}; 
finally, for
$e^-\gamma\to \nu_e H^-$ production, our code was based on that developed
in Refs.~\cite{KO,SK}.

  All processes were calculated at leading order only.
  For the SM parameters we adopted the following setup:
  $m_b=4.25$ GeV, $m_t=175$ GeV, $m_e=0.511$ MeV, $m_\tau=1.78$ GeV,
$m_\nu=0$, $M_W=80.23$ GeV, $\Gamma_W=2.08$ GeV, $M_Z=91.19$ GeV,
$\Gamma_Z=2.50$ GeV, $\sin^2\theta_W=0.232$. The top quark width $\Gamma_t$
was evaluated at leading order for each value of $M_{H^\pm}$ and
$\tan\beta$. Neutral and charged Higgs masses were calculated for given
values of $M_{A^0}$ and $\tan\beta$ using the HDECAY package \cite{hdecay},
with the SUSY masses, the trilinear couplings and the Higgsino mass
parameter $\mu$ being set to 1 TeV. The Higgs boson widths
$\Gamma_{H^\pm,\Phi^0}$ were all evaluated again by using  the
above package. In the one-loop analysis of process (\ref{loop}) we
assumed that the superpartners are sufficiently heavy to decouple, so that
only the heavy-quark loops and Higgs--gauge loops need to be included. 

We have used the energy spectrum of the back--scattered (unpolarised) photon
given by Ref.~\cite{backscattering}
\begin{equation}
F_{\gamma/e} (x)= \frac{1}{D(\xi)}\left[1-x+\frac{1}{1-x}-\frac{4x}{\xi(1-x)}
      +\frac{4x^2}{\xi^2(1-x)^2}\right],
\end{equation}
where $D(\xi)$ is the normalisation factor
\begin{equation}
D(\xi)=\left(1-\frac{4}{\xi}-\frac{8}{\xi^2}\right)\ln(1+\xi)
+\frac{1}{2}+\frac{8}{\xi}-\frac{1}{2(1+\xi)^2},
\end{equation}
and $\xi=4E_0\omega_0/m_e^2$, where
$\omega_0$ is the incoming laser photon energy
and $E_0$ the (unpolarised) positron one. In eq.~(7)
$x=\omega/E_0$ is the fraction
of the energy of the incident positron carried by the back--scattered photon,
with a maximum value
\begin{equation}\label{xmax}
x_{\mathrm {max}}=\frac{\xi}{1+\xi}.
\end{equation}
In order to maximise $\omega$ avoiding $e^+e^-$ pair creation, one takes
$\omega_0$ such that $\xi=2(1+\sqrt 2)$. So, we obtain the typical
values $\xi\simeq 4.8$, $x_{\mathrm {max}}\simeq 0.83$, $D(\xi)\simeq 1.8$,
with
$\omega_0\simeq 1.25(0.63)$ eV for a $\sqrt s_{ee} = 0.5(1)$ TeV $e^+e^-$
collider.
In the case of an $e^\pm\gamma$ scattering the total cross section $\sigma$ is
obtained by folding
the subprocess cross section $\hat\sigma$ with the photon luminosity
$F_{\gamma/e}$:
\begin{equation}
\sigma(s_{ee})=\int_{x_{\mathrm {min}}}^{x_{\mathrm {max}}}dx
F_{\gamma/e}(x)\hat\sigma(\hat s_{e\gamma}
=xs_{ee}),
\end{equation}
where $\hat s_{e\gamma}$ is the center
of mass (CM) energy at parton ($e\gamma$) level, while
\begin{equation}
x_{\mathrm {min}}=\frac {(M_{\mathrm {final}})^2}{s_{ee}},
\end{equation}
with $M_{\mathrm {final}}$ the sum of the final state particle masses.

We present the cross sections as functions of the charged Higgs boson
mass $M_{H^\pm}$ at collider energies of $\sqrt s_{ee}=
500$ and 1000 GeV and four
different values of $\tan\beta$, 1.5, 7, 30 and 40.
This is done in Fig.~\ref{rate_HpHm} for  the pair production 
process (\ref{HpHm}) --- for which there exists no $\tan\beta$ dependence,
in fact --- 
and in Figs.~\ref{rate_H0Hpm} to \ref{rate_loop} for 
the single $H^\pm$ modes. The $\tan\beta$ dependence of
processes (\ref{H0Hpm})--(\ref{loop}) can be understood as follows.
In $e^-\gamma\to\nu_e H^-\Phi^0$  ($\Phi^0=h^0, H^0, A^0$), only
couplings relevant to $W^\pm$ bosons are involved.   
If we represent the Higgs fields in the gauge basis (through a
so-called `$\beta$-rotation', i.e., a rotation 
of the mass matrix by the angle $\beta$, the ratio of
the vacuum expectation values of
the two MSSM Higgs doublets), we have two new doublets, $H_{\mathrm{SM}}$ 
(a SM-like one) and $H_{\mathrm{add}}$, 
\begin{equation}   
H_{\mathrm{add}}=\left(\matrix{ {H^+}\cr
{(\phi^0_2 + {\mathrm{i}} A^0)}/{\sqrt 2}\cr}\right),
\end{equation}
where the field $\phi^0_2$ is a superposition of the physical mass
eigenstates $h^0$ and $H^0$, and is
diagonalised through a rotation by the angle $\alpha-\beta$. 
From this formulation, it is clear that in the vertices
$W^\pm H^\mp \Phi^0$ only the cases $\Phi^0=h^0,H^0$ can carry 
a  $\tan\beta$ dependence. However, for
large $M_{H^\pm}$, one has that $\phi^0_2\to H^0$, so that such dependence
disappears to a large extent also for the case $\Phi^0=H^0$.
(In other terms, as already mentioned, in the decoupling limit
$M_{H^\pm}\to\infty$, $H^0$ carries the full gauge coupling dependence and
$h^0$ does decouple.)
Furthermore, for $e^-\gamma\to \nu_e f\bar f H^-$, 
the $\tan\beta$ dependence mainly comes from the $H^\pm f\bar f'$  
Yukawa couplings, with some minor contaminations due to $\Phi^0f\bar f$
vertices as well. There is also a resonant effect for $f=b$, in the
region $M_{H^\pm}\lsim m_t$, induced by $\bar t\to \bar b H^-$ decays (see
diagrams 2 and 6 in Fig.~\ref{feynman_graphs_ffHpm}).
Finally,  in $e^- \gamma\to \nu_e H^- $, the 
$H^\pm t\bar b$ Yukawa interaction is modulated by
the chirality structure of the loop diagrams (top-bottom
loop contributions are dominant in fact), so that
in the end the $\tan\beta$ dependence becomes  
$\sim1/\tan\beta$ or $\sim m_b^2/m_t^2\tan\beta$, 
rather than $\sim1/\tan\beta^2$ or $\sim m_b^2/m_t^2\tan\beta^2$
 (at amplitude level),
for any $M_{H^\pm}$, this explaining the enhancement for low 
$\tan\beta$ values. 

If we assume, for instance, an
integrated luminosity of 500 fb$^{-1}$
(which could be collected after a few years running 
 \cite{ginz,nlc_params}),
$10^{-5}$ pb corresponds to 5 events before acceptance cuts and
background reduction. We do not discuss the background reduction procedure
in detail in this study, and $10^{-5}$ pb is taken naively as the threshold
of the `relevance' of a process to the study of charged Higgs production
at an $e^\pm\gamma$
LC. We emphasise that this is not intended in any way as a threshold of 
detectability, or even visibility, as the evaluation of such thresholds 
would require jet simulations and machine-dependent considerations which 
are clearly beyond the scope of the current study.

The most prolific production channel is surely $e^-\gamma\to e^-
H^+H^-$, for any value of $M_{H^\pm}$ up to $\sqrt s_{ee}\approx
2 M_{H^\pm}$. However, the contribution from all other single $H^\pm$ 
production modes becomes comparable to that of the pair production mode.
In fact, after 500 inverse femtobarns
of luminosity have been collected, one
may expect between 2,000 and 110 $H^+H^-$ events to be produced,
for $M_{H^\pm}$ ranging between 140 GeV and $0.4\sqrt s_{ee}$ when
$\sqrt s_{ee}=500$ GeV, whereas corresponding numbers 
at 1000 GeV of CM energy  are 6,000 and 20. The single $H^\pm$ production
channels can altogether furnish between 
395 (122) [336] \{490\}
and 
14 (7) [12] \{17\} events
at $\sqrt s_{ee}=500$ GeV, 
corresponding to $M_{H^\pm}=140$ and 200 GeV, 
respectively, for $\tan\beta=1.5$ (7) [30] \{40\}. 
At $\sqrt s_{ee}=1000$ GeV, one instead has
877 (683) [1357] \{2201\} 
events for $M_{H^\pm}=140$ GeV
and
15 (2) [8] \{13\} for $M_{H^\pm}=400$ GeV.

Above the kinematic threshold of pair production, i.e.,
when $2M_{H^\pm}\gsim 0.8\sqrt s_{ee}$, only the loop-mediated 
process $e^-\gamma\to \nu_e H^-$ can in principle be useful, at least at
low $\tan\beta$.
In fact, for $\tan\beta=1.5$, one has 10 
events when $\sqrt s_{ee}=500$ GeV and $M_{H^\pm}=300$ GeV,
or $\sqrt s_{ee}=1000$ GeV and $M_{H^\pm}=600$ GeV. For such heavy masses, 
all other single $H^\pm$ channels become negligible,
even  at large $\tan\beta$.
Further notice the much steeper descent in the production
rates of processes (\ref{H0Hpm})--(\ref{ffHpm}), 
Figs.~\ref{rate_H0Hpm}--\ref{rate_ffHpm}, with respect to those
of process (\ref{loop}), see Figs.~\ref{rate_loop}, with
growing Higgs mass values.
 
 \section{Possible signals and detection strategies}

Over most of the heavy mass range, $M_{H^\pm}\ge m_t$, charged Higgs bosons
decay to $t\bar b$ (and charge conjugate) pairs
\cite{BRs}. Given the not so large production rates in all modes considered,
it is natural to focus on this decay channel
first\footnote{In the pair production
mode, one may alternatively conceive to ask for one of the
two charged Higgs bosons to decay via $H^-\to\tau\nu_\tau$, assuming
large $\tan\beta$ values, where the corresponding Branching Ratio (BR)
can be as large as 10\% \cite{BRs}. However, we do not consider
here this possibility.}.

If both charged Higgs bosons decay to top-bottom pairs, the final signature
produced by the $H^+H^-$ production channel is 
$b\bar b b\bar bW^+W^-$\footnote{The $H^+H^-\to t\bar b \bar t b$ decay
combination can be complemented by the $H^+H^-\to W^+h^0 W^- h^0$ one,
at low $\tan\beta$, and mixtures of the four $H^\pm$ decays modes, all
producing the $b\bar b b\bar bW^+W^-$ intermediate stage.}, with
the very forward/backward electron escaping detection.
Under these circumstances, one may apply the same selection procedure
outlined in Ref.~\cite{Marco2}, for the
$e^+e^-\to H^+H^-$ case. Here,
the two $W^\pm$'s are required to decay hadronically, hence, the
final signal is made up by eight jets. Four of these can be $b$-tagged
with high efficiency. Both $W^\pm$'s and $t$'s are reconstructed.
After completing the jet assignment, one can finally perform a kinematical
fit, imposing the constraint of equal Higgs boson masses. This way, the
signal should clearly emerge from the background with high statistical
significance. The latter is expected to mainly be constituted
by $e^-\gamma \to e^- t\bar t h^0$ events, with $h^0\to b\bar b$,
the counterpart of $e^+e^-\to t\bar h^0\to t\bar t b\bar b$, 
discussed in the above paper. Another noise could be induced in the
$e^\pm\gamma$ case
by triple-gauge-vector production, via $e^-\gamma\to e^- Z^0 W^+W^-$,
with $W^+W^-\to jjjj$ and $Z\to jjjj$ too. This background
can however be suppressed by imposing $M_{jjjj}\ne M_{Z^0}$, even
before enforcing $b$-tagging.  (Events of the type 
$e^-\gamma\to \nu_e W^-W^+W^-$, with
a longitudinal $W^-$ boson exchange, could in principle be relevant;
in practise to impose $b$-tagging should be enough to 
remove them efficiently.) 

The signature expected from the single $H^\pm$ production modes
(\ref{H0Hpm})--(\ref{ffHpm}) is either the one above, or else
 $b\bar b b\bar b W^-$, with $W^-\to jj$, yielding one less jet pair 
(again, assuming $H^-\to \bar t b$ hadronic decays). 
On the one hand, it should be noticed that
processes of the type (\ref{ffHpm}) with $f=\tau,\nu_\tau$ are
smaller in comparison to the case $f=b$. On the other hand, for 
$M_{\Phi^0}$ below 400 GeV, the dominant decays of neutral Higgs
bosons are either $\Phi^0\to b\bar b$ (yielding the six-jet signature)
or $H^0\to h^0h^0, W^+W^-,Z^0Z^0$ and $A^0\to Z^0h^0$, with 
$h^0\to b\bar b$ and $W^\pm,Z^0\to jj$ (yielding the eight-jet signature).
Thus, one could pursue in either case a selection strategy along the lines
already described, with the only caution of constraining the final
kinematical fit to different intermediate masses ($M_{h^0}$, 
$M_{W^\pm}$ and/or $M_{Z^0}$, rather than a second $m_t$). Unfortunately,
in the single $H^\pm$ production cases, one can no longer impose
the equal Higgs mass constraints, which revealed itself rather
effective in rejecting both combinatorial and genuine background
in the $e^+e^-\to H^+H^-$ case. Significant noises in the eight-jet
channel are as above, whereas in the six-jet case one may expect
$e^-\gamma\to e^- t\bar t$ (again, the electron is assumed to be
undetected), with $t\bar t\to b\bar b W^+W^-$
and $W^+W^-$ yielding in turn four jets, to be relevant. This can however 
be suppressed  by a triple (at least) $b$-tagging and/or a mass rejection, 
$M_{jj}\ne M_{W^\pm}$, against a second jet-pair reconstructing a $W^\pm$
mass.

The total production rate for the  $\nu_e b\bar b H^-$ final state
can be found in Figs.~\ref{rate_bbHpm} (upper curves). These
have been computed by adding to the diagrams in 
Fig.~\ref{feynman_graphs_ffHpm} (for the case $f=b$ and $f'=t$)
those in Fig.~\ref{feynman_graphs_H0Hpm}, the latter supplemented
by the decay currents $\Phi^0\to b\bar b$ (including Higgs propagator
effects), then taking the square of
the sum of all such diagrams.
This way, interference effects between the two channels are
taken into account appropriately. In the total cross sections,
one may appreciate the different components of $b\bar b H^-$ final states.
When $M_{H^\pm}\lsim m_t$, there is a resonant contribution from
$\bar t \to \bar b H^-$ decays (graphs 2,5,7 and 11
in Fig.~\ref{feynman_graphs_ffHpm}), which is clearly 
visible in the lower curves. The resonant $\Phi^0\to b\bar b$ contributions are
responsible for the general increase of the production rates for
$M_{H^\pm}\lsim \sqrt s_{e\gamma}/2$ (upper curves versus lowers curves
in Figs.~\ref{rate_bbHpm}). The relative strength of the two
resonant contributions in the allowed kinematic regions
is regulated by $\tan\beta$ \cite{BRs}. The contributions
from the Higgs-strahlung diagrams, namely 3 and 8 in
Fig.~\ref{feynman_graphs_ffHpm}, is much smaller at both energies
considered, and can yield rates in small excess of ${\cal O}(10^{-5})$
fb only at $\tan\beta=40$ and $M_{H^\pm}\lsim m_t$.
 
Process (\ref{loop}) has already been studied in Ref.~\cite{KO}, where
a detailed signal-to-background analysis has been carried out,
for the dominant charged Higgs decay channel, to top-bottom pairs,
with respect the continuum production $e^-\gamma \to \nu_e \bar t b$.
Unfortunately, despite the signal is above our threshold
of relevance up to very large masses, well beyond $\sqrt s_{e\gamma}/2$
(as already remarked),
for both CM energy considered and at small  $\tan\beta$ (see 
Fig.~\ref{rate_loop}), the mentioned irreducible noise 
is in the end prohibitive (besides, notice that the
latter scales with $\tan\beta$ exactly as the signal does, 
see Ref.~\cite{KO}). Finally, it was also pointed out in the
study of Ref.~\cite{KO})
 the negative interference effects between signal and background,
which further deplete the signal-to-background rates.

To summarise our findings, the interesting production modes
(\ref{HpHm})--(\ref{loop}), which are specific to the $e^\pm\gamma$ option, 
have cross sections which
are too small to be of much use, and the quantities which
can in principle be measured there (such as multiple Higgs-gauge-boson
and Higgs-fermion couplings) can be better accessed in  $e^+e^-$
collisions, as can be confirmed by comparing the results 
presented here with those obtained in Ref.~\cite{epem}.

 \section{Conclusions}

We have verified that the potential of future LCs operating in the
$e^\pm\gamma$ mode in covering the heavy charged Higgs boson sector of
the MSSM is only limited to values of the charged Higgs boson
mass compatible with the kinematic constraint $2M_{H^\pm}\lsim \sqrt
s_{e\gamma} \approx 0.8\sqrt
s_{ee}$. However, over this range, not only the pair production mode
 $e^-\gamma \to e^- H^+H^-$ is large, but also a variety of
single $H^\pm$ production channels, $e^-\gamma\to \nu_e H^- \Phi^0 $
(when $\Phi^0=H^0$ and $A^0$), $e^-\gamma\to \nu_e b\bar b H^-$ and
$e^-\gamma\to \nu_e H^-$, can produce rates of the same order of magnitude. 
In general, we have so far established that the $e^+e^-$ beam option of
future LCs offers better chances than the $e^\pm\gamma$ one of
detecting and studying heavy charged Higgs bosons of the MSSM 
(recall Ref.~\cite{epem}).
In fact, the arguments adopted here and in Ref.~\cite{epem} can
equally be applied to a more general Two-Higgs-Doublet-Model (2HDM).

The drawbacks of the $e^\pm\gamma$ beam option in comparison
to the $e^+e^-$ are twofold. Firstly, Higgs cross sections
are significantly smaller, both for pair production and single $H^\pm$
channels. Secondly, when $M_{H^\pm}$ is significantly larger
than $\sqrt s_{e\gamma}/2$ --- so that $H^+H^-$ final states
are no longer available --- none of the
single production modes considered here is able to furnish enough
events to pursue a statistically significant analysis. 
 Not even the loop-induced mode
$e^- \gamma\rightarrow \nu_e H^-$ is very helpful.
Here, despite the fact that the production rates
are very stable up to very large $M_{H^\pm}$ values
(indeed, comparable to $\sqrt s_{e\gamma}$) 
and that in principle the signal could
be observable at very low $\tan\beta$ (in which case though,
one should dismiss the MSSM in favour of a general 2HDM), 
one has to cope with a large irreducible background in non-resonant
$e^- \gamma\to \nu_e \bar t b$ events, overwhelming the signal in the
$H^-\to \bar t b$ decay mode, even before taking into account the 
negative interference between the two competing processes \cite{KO}.
This situation is in contrast with the case of single charged Higgs
production at the $e^+e^-$ option, where it has been revealed
that there are several channels which are viable
complements to the pair production mode \cite{epem}.

Finally, although at
the energy scales which we have considered in this work the
$e^\pm\gamma$ option offers no advantage compared to the $e^+e^-$ one,
the situation would get somewhat brighter at higher CM energies.
We note that the cross sections typically behave as $1/s_{ee}$ in annihilation
reactions (i.e., in the $e^+e^-$ option of a LC), whereas the 
$t$-channel induced $e^\pm\gamma$
 processes have cross sections which scale as $\log(s_{e\gamma})$ with
increasing energy. At 5 TeV \cite{CLIC}, for example, we believe
that the $e^\pm\gamma$ option would offer physics opportunities
that can complement those available
in $e^+e^-$ and $\gamma\gamma$ LCs.  However, we have
not pursued here this possibility, as we have confined ourselves to
current design values of $\sqrt s_{ee}$ at TESLA
\cite{TESLA}, i.e., in the  TeV region.

A similar analysis \cite{gamgam}
to the one performed here and in Ref.~\cite{epem}
is now in progress for the $\gamma\gamma$ beam option of future LCs.

 \subsection*{Acknowledgements}

SK would like to thank C.-P. Yuan for useful discussions. 
SK and KO would like to thank KEK for enabling KO's
visit to Osaka and for hospitality during SK's
visit to KEK. SK and KO also thank Y. Okada
for discussions.

\newpage

  \FIGURE[p]{
  \hskip-3.0cm
  \epsfig{file=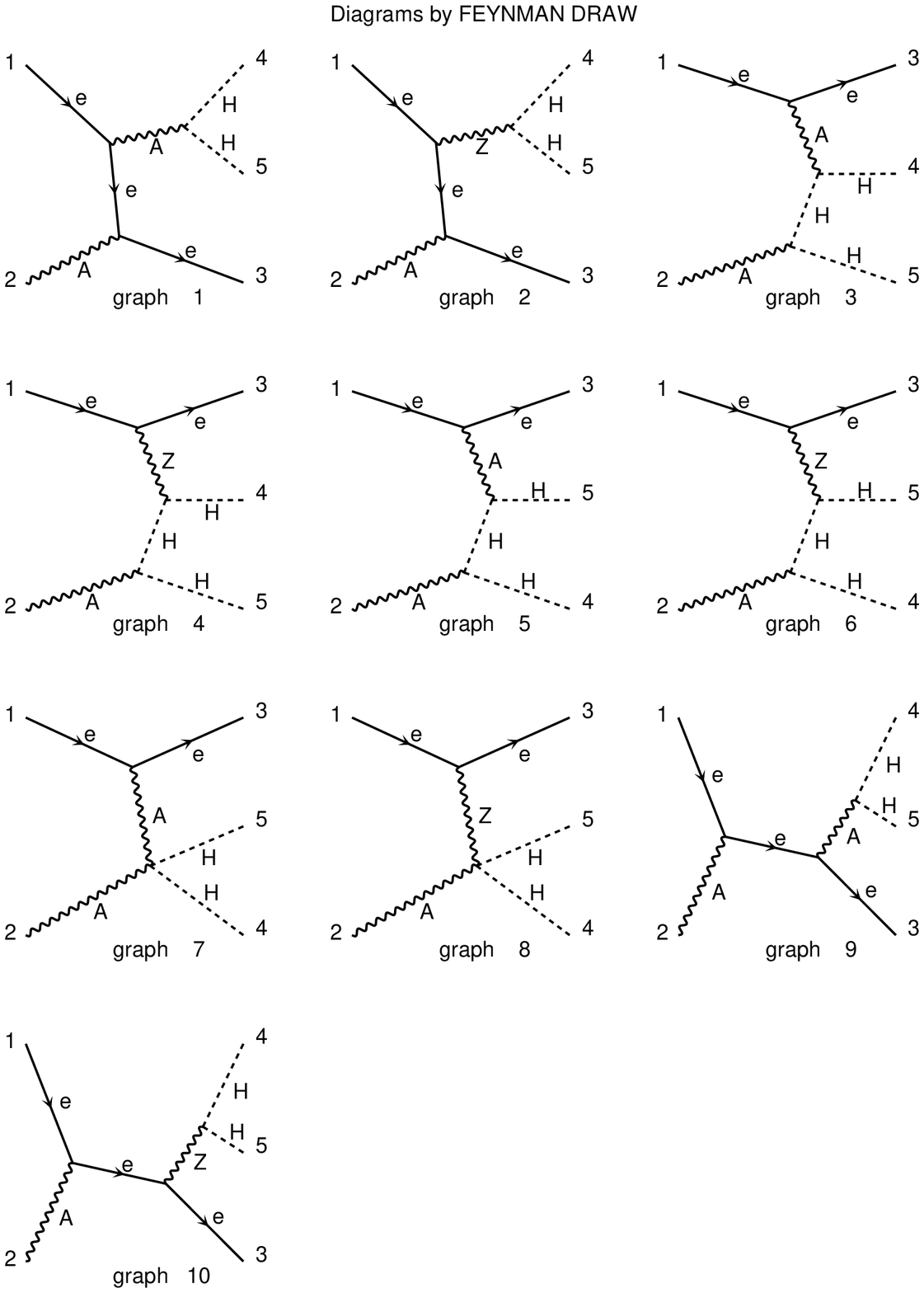,width=12cm,angle=0}\\
  \vspace{-4.0truecm}
  \caption{Feynman diagrams for process (\ref{HpHm}).
  The labels {\tt e, A, Z and H} refer
  to an electron, $\gamma$, $Z$ and to both neutral
  and charged Higgs bosons, as appropriate, respectively.}
  \label{feynman_graphs_HpHm}
  }

  \FIGURE[p]{
  \hskip-3.0cm
  \epsfig{file=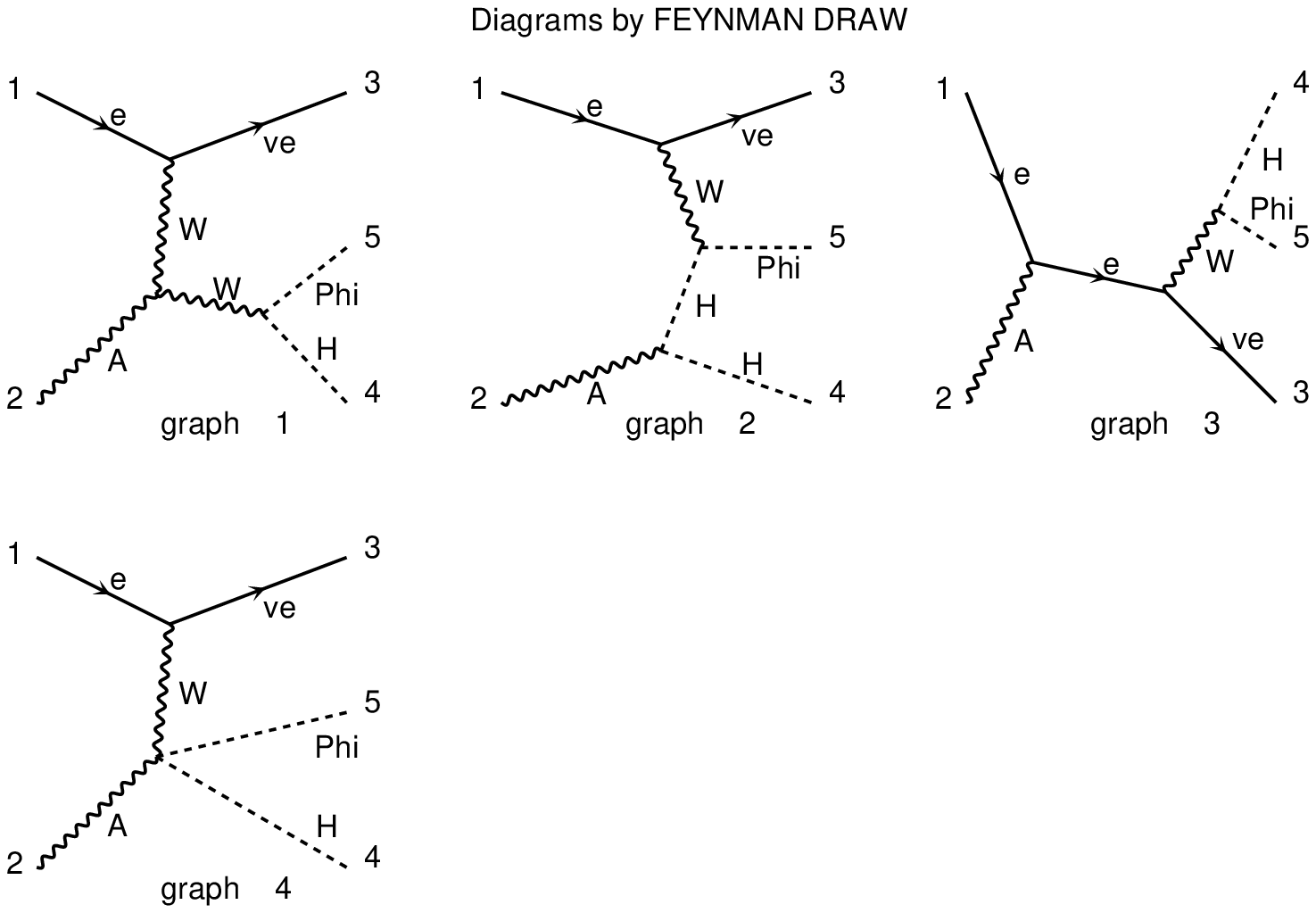,width=12cm,angle=0}\\
  \vspace{-12.0truecm}
  \caption{Feynman diagrams for processes of the type (\ref{H0Hpm}).
  The labels {\tt e, ve, A, W and H(Phi)} refer
  to an electron, neutrino, $\gamma$, $W^\pm$ and 
  a charged(neutral) Higgs boson, respectively.}
  \label{feynman_graphs_H0Hpm}
  }

  \FIGURE[p]{
  \hskip-3.0cm
  \epsfig{file=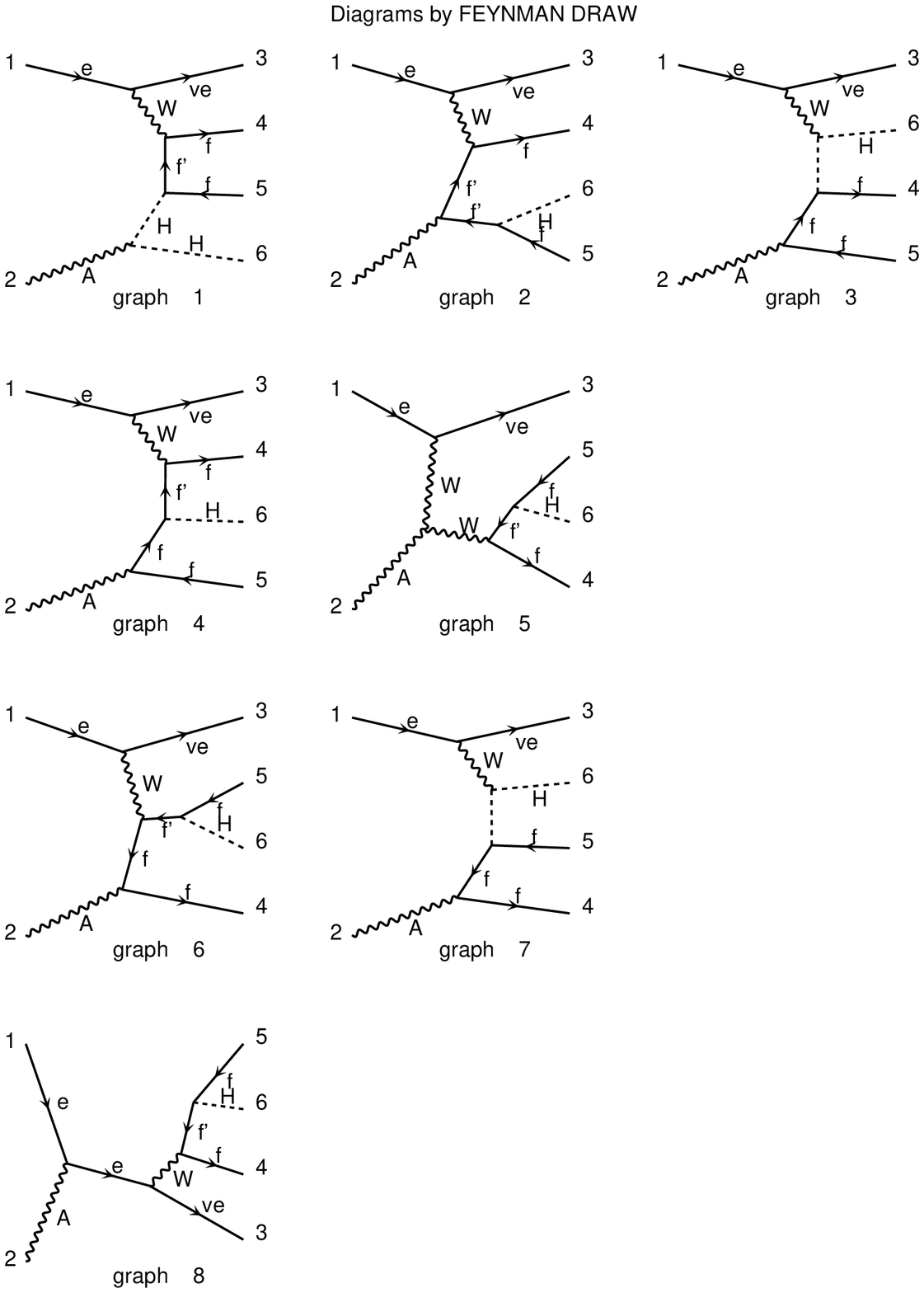,width=12cm,angle=0}\\
  \vspace{-5.0truecm}
  \caption{Feynman diagrams for processes of the type (\ref{ffHpm}).
  The labels {\tt e, ve, A, W and H} refer
  to an electron, neutrino, $\gamma$, $W^\pm$ and 
  a charged Higgs boson, respectively, whereas $f'$ and $f$
  refer to $b$- and $t$-quarks or $\tau$- and $\nu_\tau$-leptons,
  as appropriate.}
  \label{feynman_graphs_ffHpm}
  }

  \FIGURE[p]{
  \hskip-3.0cm
  \epsfig{file=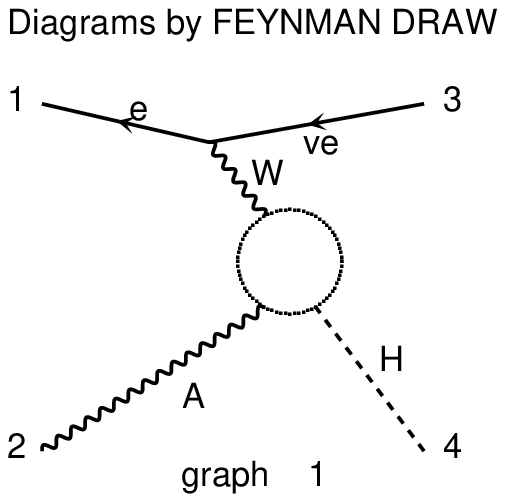,width=12cm,angle=0}\\
  \vspace{-16.0truecm}
  \caption{Feynman diagrams for processes of the type (\ref{loop}).
  Labels are as in Fig.~\ref{feynman_graphs_H0Hpm} (apart
  from {\tt Phi}). The loop particle content is detailed in Ref.~\cite{SK}.}
  \label{feynman_graphs_loop}
  }

  \FIGURE[p]{
  \epsfig{file=egamehphm.ps,width=8cm,angle=90}\\
  \caption{Total cross sections for process (\ref{HpHm}).}
  \label{rate_HpHm}
  }

  \FIGURE[p]{
  \epsfig{file=egamnhpmh02_500.ps ,width=5cm,angle=90}
  \epsfig{file=egamnhpmh02_1000.ps,width=5cm,angle=90}\\
  {\small 
  (a) Process $e^-\gamma\to \nu_e \Phi^0 H^-$ with $\Phi^0=h^0$.}\\[1cm]
  \epsfig{file=egamnhpmh01_500.ps ,width=5cm,angle=90}
  \epsfig{file=egamnhpmh01_1000.ps,width=5cm,angle=90}\\
  {\small 
  (b) Process $e^-\gamma\to \nu_e \Phi^0 H^-$ with $\Phi^0=H^0$.}\\[1cm]
  \epsfig{file=egamnhpmh03_500.ps ,width=5cm,angle=90}
  \epsfig{file=egamnhpmh03_1000.ps,width=5cm,angle=90}\\
  {\small 
  (c) Process $e^-\gamma\to \nu_e \Phi^0 H^-$ with $\Phi^0=A^0$.}\\[1cm]
  \caption{Total cross sections for processes of the type (\ref{H0Hpm}).
  In (c), the four curves in each plot coincide within graphical
  resolution.}
  \label{rate_H0Hpm}
  }

  \FIGURE[p]{
  \epsfig{file=egamnbbhpm_500.ps ,width=5cm,angle=90}
  \epsfig{file=egamnbbhpm_1000.ps,width=5cm,angle=90}\\
  {\small 
  (a) Process $e^-\gamma\to \nu_e f\bar f H^-$ with $f=b$.}\\[1cm]
  \epsfig{file=egamntatahpm_500.ps ,width=5cm,angle=90}
  \epsfig{file=egamntatahpm_1000.ps,width=5cm,angle=90}\\
  {\small 
  (b) Process $e^-\gamma\to \nu_e f\bar f  H^-$ with $f=\tau$.}\\[1cm]
  \epsfig{file=egamnvtvthpm_500.ps ,width=5cm,angle=90}
  \epsfig{file=egamnvtvthpm_1000.ps,width=5cm,angle=90}\\
  {\small 
  (c) Process $e^-\gamma\to \nu_e f\bar f  H^-$ with $f=\nu_\tau$.}\\[1cm]
  \caption{Total cross sections for processes of the type (\ref{ffHpm}).}
  \label{rate_ffHpm}
  }

  \FIGURE[p]{
  \epsfig{file=egamnhpm_500.ps ,width=5cm,angle=90}
  \epsfig{file=egamnhpm_1000.ps,width=5cm,angle=90}\\
  \caption{Total cross sections for process (\ref{loop}).}
  \label{rate_loop}
  }

  \FIGURE[p]{
  \epsfig{file=res_egamnbbhpm_500.ps ,width=5cm,angle=90}
  \epsfig{file=res_egamnbbhpm_1000.ps,width=5cm,angle=90}\\
  \caption{Total cross sections for processes of the type (\ref{H0Hpm}),
  with $\Phi^0\to b\bar b$, plus  those of the type  (\ref{ffHpm}),
  with $f,f'=b,t$, including the interference (upper
  curves), compared to the latter only (lower curves).}
  \label{rate_bbHpm}
  }

 \end{document}